\documentclass[twocolumn,PRB,amsmath,amssymb,showpacs]{revtex4}
\usepackage{amssymb}
\usepackage{amsmath}
\usepackage{graphicx}
\usepackage{epstopdf}
\usepackage{bm}%
\usepackage[normalem]{ulem}
\usepackage{epsfig}
\usepackage{graphicx}
\usepackage{amsfonts,amssymb}
\usepackage{amsmath}
\usepackage{color}
\usepackage {times}
\usepackage{epstopdf}
\usepackage [utf8]{inputenc}

\newcommand{\bl}{\begin{aligned}}
\newcommand{\el}{\end{aligned}}

\newcommand{\be}{\begin{equation}}
\newcommand{\ee}{\end{equation}}   
\newcommand{\bea}{\begin{eqnarray}}
\newcommand{\eea}{\end{eqnarray}}
\newcommand{\ba}{\begin{array}}
\newcommand{\ea}{\end{array}}

\newcommand{\q}{{\bf q}}
\renewcommand{\k}{{\bf k}}
\newcommand{\Q}{{\bf Q}}

\graphicspath{{Figs/}}

\begin{document}

\title{Effect of strain-induced orbital splitting on the magnetic excitations in undoped cuprates}

\date{\today}
\author{Dheeraj Kumar Singh$^{1}$}
\author{Yunkyu Bang$^{2,3}$}
\affiliation{$^1$School of Physics and Materials Science, Thapar Institute of Engineering 
and Technology, Patiala-147004, Punjab, India}
\affiliation{$^2$Department of Physics, POSTECH, Pohang 790-784, Korea}
\affiliation{$^3$Asia Pacific Center for Theoretical Physics, Pohang, Gyeongbuk 790-784, Korea}
\begin{abstract}
We investigate the magnetic excitations in view of the recent reports 
suggesting that the spin-wave energy may exhibit a significant dependence on the in-plane 
strain of a thin film of La$_2$CuO$_4$. The nature of dependence, as we find, can be 
explained naturally within a two-orbital model based on the $d_{x^2-y^2}$ and $d_{3z^2-r^2}$ orbitals. 
In particular, as the orbital-splitting energy between the $d_{x^2-y^2}$ and $d_{3z^2-r^2}$ orbitals 
increases with compressive strain, the zone-boundary spin-wave energy hardens. However, the hardening 
persists only until the orbital splitting reaches $\sim$ 2eV, beyond which there is no significant change. The 
behavior of zone-boundary spin-wave energy is explained in terms of the 
extent of hybridization between one of the exchange-split $d_{x^2-y^2}$ band which is 
nearly half filled and the $d_{3z^2-r^2}$ band. The role of second-order antiferromagnetic superexchange
process involving the inter-orbital hopping is also discussed.
\end{abstract}
\maketitle
\newpage
\section{introduction}
The origin of unconventional superconductivity has been a recurrent theme since the discovery of 
high-$T_c$ cuprates in the late eighties~\cite{bednorz,imada,damacelli,lee}. The last decade has witnessed the discovery 
of another large family of multiband superconductors based on iron, which are also believed widely to be 
unconventional in nature~\cite{kamihara,boeri}. A striking similarity between the two class of superconductors is that a long-range
magnetic order is exhibited by the parent compounds that gives way to superconductivity on doping either holes or electrons~\cite{si}. Thus, 
the idea that the unconventional superconductivity may be mediated by the spin fluctuations is strengthened 
further and therefore the nature of such fluctuations can be the key to the understanding of pairing mechanism.

The spin-wave excitations in the Mott-antiferromagnetic phase of high-$T_c$ cuprates show a more dispersive 
behavior near the zone-boundary in comparison to the Heisenberg antiferromagnet with
only nearest-neighbor exchange coupling~\cite{coldea,braicovich,headings,dean,peng}. 
The deviation was explained by incorporating the exchange couplings beyond the nearest neighbor in the 
Heisenberg model or by considering hopping beyond the nearest neighbor in the one-orbital Hubbard 
model~\cite{avinash,delannoy,carmelo}. Recent experiments based mainly 
on the resonant inelastic x-ray spectroscopy (RIXS) have unfolded several new features which are difficult 
to explain within the one-orbital model~\cite{ivashko1,ivashko2}. One such remarkable feature is that the
spin-wave energy exhibits a variation of $\sim$ 60meV at the zone-boundary upon
subjecting a thin film of cuprate to a substrate-induced strain. For instance, the spin-wave energy shows hardening near the 
zone boundary with growing in-plain compressive strain.

In the presence of in-plane strain, the orbital overlap~\cite{abrecht, ivashko2} and on-site Coulombic repulsion
~\cite{abrecht,tomczak1, tomszak2,kim} can get affected. The compressive strain enhances the orbital overlap,
which results into an increase in the in-plane hopping parameters ($t$). On the other hand, the separation between the
two $e_g$ levels also grows, which is expected to push the $d_{3z^2-r^2}$-band further below the Fermi level so that the screening of 
the intra-orbital Coulombic interaction ($U$) for $d_{x^2-y^2}$ orbital gets reduced resulting into an increase in 
$U$. However, $U/t$ may remain constant as 
suggested by a density-functional theory (DFT) calculation and x-ray absorption spectrum (XAS) measurement~\cite{ivashko2}. 
Consequently, the effective exchange coupling $J \approx 4t^2/U$ can increase
in the limit of a very large $U$, which has been linked to the hardening
of zone-boundary spin-wave energy. Since the separation between the two $e_g$ orbitals is directly affected by the 
in-plane strain, a study based on a model incorporating 
$d_{3z^2-r^2}$-orbital can provide a more
clear picture about the origin of variation of zone-boundary spin-wave 
energy with orbital-splitting (OS), which is undertaken in the current paper.

The importance of $e_g$ OS has been emphasized in several recent works including the one which suggested 
that the difference between the superconducting transition temperature across the 
high-$T_c$ cuprates may depend on the $e_g$-level separation. Particularly, the superconducting transition 
temperature was shown to increase with OS~\cite{sakakibara,sakakibara1,sakakibara2,tang}. The $e_g$ OS ($\delta$) 
can range in between 1eV $\lesssim$ $\delta$ $\lesssim$ 2eV, whereas the splitting 
between the two sets $e_g$ and $t_{2g}$ of orbitals is  $\approx$ 2eV~\cite{lorenzana,imada,
graaf,hill,sala,ghiringhelli,hozoi,jang}. Therefore, while the 
$d_{x^2-y^2}$ orbital based one-orbital model can describe the correlation effects for cuprates
with a larger $e_g$ splitting, it becomes necessary to include both the $e_g$ orbitals for the cuprates with a
smaller splitting. The evidence from the angle-resolved photoelectron 
spectroscopy (ARPES) experiments indicates a significant hybridization of bands located 
not far from the Fermi level, which involves $d_{3z^2-r^2}$ orbital ~\cite{matt,kramer}. An 
important role of $d_{3z^2-r^2}$ orbital was also indicated in a recent work examining the spin-wave 
excitations in the hole-doped La$_2$CuO$_4$ (LSCO)~\cite{ivashko1}. Presence of this additional orbital may also be responsible for 
the stability of AFM state against hole doping resulting mainly from the Hund's first rule which demands 
the maximization of total spin \cite{dheeraj}.

In this paper, we investigate the role of OS between $d_{x^2-y^2}$ and $d_{3z^2-r^2}$ orbitals in  
the spin-wave excitations for the AFM phase of the undoped cuprate. In order to achieve this goal, we consider a two-orbital 
model based on both the $e_g$ orbitals. Our findings indicate that (i) the zone-boundary spin-wave energy increases 
with the in-plane compressive strain in the cuprates with a relatively smaller $e_g$ splitting such as
LSCO, a result in qualitative agreement with recent RIXS experiment. (ii) However, it does not show
any significant dependence on strain
for the cuprates with a larger splitting. (iii) The behavior originates from
the orbital mixing of the lower exchange-split 
$d_{x^2-y^2}$ band and a nearly flat $d_{3z^2-r^2}$ band. This mixing generates additional exchange coupling based on 
a second order inter-orbital superexchange process in addition to the intra-orbital superexchange.
\section{Model}
We consider a Hamiltonian based on the two $e_g$ orbitals. The delocalization-energy 
gain term is given by 
\begin{equation}
 \mathcal{H}_{KE}  = \sum_{{\bf i}{\bf j}}\sum_{\mu,\nu,\sigma}
 t_{{\bf i}{\bf j}}^{\mu \nu} 
d_{{\bf i} \mu \sigma}^\dagger d_{{\bf j} \nu \sigma}^{}.
\end{equation}
$t_{{\bf i}{\bf j}}^{\mu \nu}$s are the hopping matrix elements
from the orbital $\mu$ at site ${\bf i}$ to the orbital $\nu$ at site ${\bf j}$, respectively. 
The operator $d_{{\bf i} \mu \sigma}^\dagger$ ($d_{{\bf i} \mu \sigma}$) creates
(destroys) an electron with spin $\sigma$ at site ${\bf i}$ in the orbital
${\mu}$. The orbitals $\mu$ and $\nu$ are either of the Cu 3$d_{x^2-y^2}$ and $d_{3z^2-r^2}$ Wannier orbitals. 
The $d_{x^2-y^2}$ Wannier orbital arises due to 3$d_{x^2-y^2}$ orbital of Cu and the bridging $2p_{x/y}$ orbital of O 
located in between two Cu atoms in the CuO$_2$ plane. $d_{3z^2-r^2}$ Wannier orbital results from the 
Cu $d_{3z^2-r^2}$ orbitals and $2p_z$ orbital of O present in the apical position~\cite{sakakibara1}.

The orbital splitting between
the $e_g$ orbitals $d_{x^2-y^2}$ and $d_{3z^2-r^2}$ is given by 
\be
\mathcal{H}_{OS}  =
\frac{\delta}{2}\sum_{\bf i} 
(d_{{\bf i} \gamma \sigma}^\dagger d_{{\bf i} \gamma \sigma}^{}- d_{{\bf i} \gamma^{\prime}
\sigma}^\dagger d_{{\bf i}\gamma^{\prime} \sigma}),
\ee
where $\gamma$ and $\gamma^{\prime}$ denote $d_{x^2-y^2}$ and $d_{3z^2-r^2}$ orbitals, respectively. 
$\delta$ is the orbital splitting parameter, which
is controlled by the distance of the apical oxygen from the CuO$_2$ plane. 
The in-plane strain applied on a thin film can generate a modification in both in-plane and
out-of-plane lattice parameter, which can introduce a change in the overall crystal-field effect. 
Consequently, $\delta$ gets directly affected. The same has been indicated by the XAS and RIXS measurements. The $dd$ excitations 
study based on the Cu $L_3$ edge, shows that the center of mass of $dd$ excitations shift systematically towards 
higher energy with increasing in-plane compressive strain. The dependence of the position of 
the center of mass on the strain parameter defined as 
$\epsilon = (a - a_0)/a_0$ is nearly linear. This may also imply a similar enhancement in the $e_g$ orbital 
splittings as a function of $\epsilon$~\cite{ivashko2}.

The standard on-site Coulomb interaction is given by
\bea
\mathcal{H}_{int} 
&=& U \sum_{{\bf i},\mu} n_{{\bf i} \mu \uparrow} 
n_{{\bf i} \mu \downarrow} 
+ (U' -
\frac{J}{2}) \sum_{{\bf i}} n_{{\bf i} \gamma} n_{{\bf i} \gamma^{\prime}}  \nonumber\\
&-& 2 J \sum_{{\bf i}} {\bf S}_{{\bf i} \gamma} \cdot {\bf S}_{{\bf i} \gamma^{\prime}}  + J \sum_{{\bf i}, \sigma} 
d_{{\bf i} \gamma \sigma}^{\dagger}d_{{\bf i} \gamma \bar{\sigma}}^{\dagger}d_{{\bf i} \gamma^{\prime} \bar{\sigma}}^{}
d_{{\bf i} \gamma^{\prime} \sigma}^{}.
\label{int}
\eea 
The intra- and inter-orbital Coulomb interaction ($U$ and $U^{\prime}$) terms are 
described by the first and second terms, respectively, where $n_{{\bf i} \mu \sigma} =  d^{\dagger}_
{{\bf i} \mu \sigma}d^{}_{{\bf i} \mu \sigma}$ and $n_{{\bf i} \gamma} =  \sum_{\sigma} d^{\dagger}_
{{\bf i} \gamma \sigma}d^{}_{{\bf i} \gamma \sigma}$ with $\sigma = \uparrow, \downarrow$. The third term
stands for the Hund's coupling between
electrons of different orbitals, where ${ S}^{l}_{{\bf i} \gamma} =  \sum_{\sigma \sigma^{\prime}} d^{\dagger}_
{{\bf i} \gamma \sigma} {\sigma}^l_{\sigma \sigma^{\prime}} d^{}_{{\bf i} \gamma \sigma}$.${\sigma}^ls$ are the Pauli matrices with $l = x, y, z$. The last term is 
the pair-hopping term.
\begin{figure}[]
\begin{center}
\vspace{-2mm}
\psfig{figure=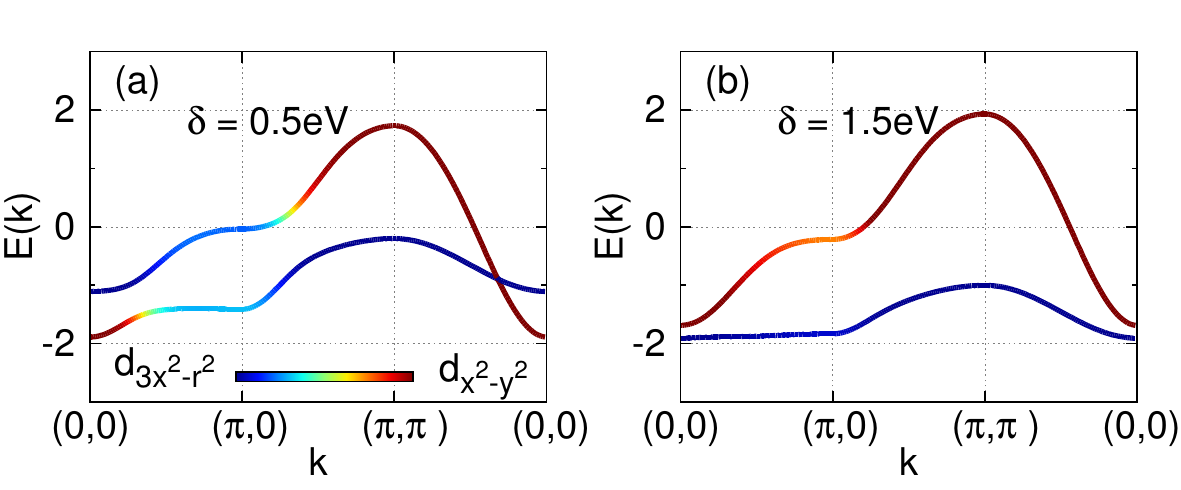,width=1.0 \linewidth,angle=0}
\end{center}
\vspace{-6mm}
\caption{The orbital contents of electronic bands along the high-symmetry directions in the two-orbital model when the 
orbital splittings ($\delta$) are (a) 0.5eV and (b) 1.5eV. The orbital mixing is dominant near $(\pi, 0)$.}
\vspace{-4mm}
\label{band}
\end{figure}

\section{Method}

The mean-field decoupling of various interaction terms in Eq. 1 originating 
from the Coulombic interaction yields the following mean-field Hamiltonian~\cite{kovacic}
\be
{{H}}_{{}\k} = 
\sum_{\k \sigma}\Psi^{\dagger}_{{\bf k} \sigma}
\begin{bmatrix}
 \hat{ h}({\k})+\hat{N} \,& \,{\rm sgn}\bar{\sigma}\hat{M} \\
 {\rm sgn}\bar{\sigma}\hat{M} \,& \,\hat{ h}({\bf {k+Q}})+\hat{N}
\end{bmatrix} 
\Psi_{{\bf k} \sigma},
\ee
for the ($\pi, \pi$) AFM state in the momentum space. $\Psi^{\dagger}_{{\bf k} \sigma} = (d^{\dagger}_{{\k}1\sigma}, d^{\dagger}_{{\k}2\sigma},
{d}^{\dagger}_{{\k}\bar{1}\sigma},{d}^{\dagger}_{\k\bar{2}\sigma})$  with
 ${d}^{\dagger}_{{\k}\bar{l}\sigma} = d^{\dagger}_{{\k+\Q}l\sigma}$
 and $\Q = (\pi, \pi)$. The elements of the 2$\times$2 matrix $\hat{ h}_{\k}$ are given by 
 \bea
 h_{11} (\k) &=& -2t_1 (\cos k_x + \cos k_y) + 4 t_2 \cos k_x \cos k_y \nonumber\\
     &-& 2 t_3 (\cos 2k_x + \cos 2k_y)  \nonumber\\
h_{12} (\k) &=& h_{21}(\k) =  2 t_4 (\cos k_x - \cos k_y) \nonumber\\
&+& 2 t_5 (\cos 2k_x - \cos 2k_y) \nonumber\\
h_{22} (\k) &=& -2 t_6 (\cos k_x + \cos k_y),
 \eea
where the hopping parameters are $t_1 = 0.452$, $t_2 = 0.0895$, $t_3 = 0.0705$,
$t_4 = 0.171$, $t_5 = 0.0248$, $t_6 = 0.113$ with the unit being eV. $t_1$, $t_2$ and $t_3$ are the nearest, next-nearest and 
next-next-nearest neighbor intra-orbital hopping parameters for the $d_{x^2-y^2}$ orbitals. $t_4$ 
and $t_5$ are the nearest and next-next-nearest neighbor inter-orbital hopping parameters. $t_6$ is the 
nearest neighbor intra-orbital hopping parameter for the $d_{3z^2-r^2}$ orbitals.
 
$\hat{M}$ and $\hat{N}$ are 2$\times$2 matrices with 
the elements given in terms of the interaction parameters, charge densities and magnetization. 
{$2M_{ll} = Um_{ll}+J\sum_{l \ne m}m_{mm}$ and $2M_{lm} = Jm_{lm}+(U-2J)m_{ml}$. 
Also, $2N_{ll} = Un_{ll}+(2U-5J)\sum_{l \ne m}n_{mm}$ and $2N_{lm} = Jn_{lm}+(4J-U)n_{ml}$. The self-consistent mean-field 
order parameters, \textit{i.e.} charge density and magnetization are given by 
$n_{\mu \nu} = \sum_{\k \sigma} \langle d^{\dagger}_{\k \mu \sigma}d^{}_{\k \nu \sigma}\rangle$ and 
$m_{\mu \nu} = \sum_{\k \sigma} \langle d^{\dagger}_{\k+\Q {\mu} \sigma}d^{}_{\k \nu \sigma}\rangle {\rm sgn}\sigma$.

In order to study the spin-wave excitations in the AFM state, we 
calculate the transverse spin susceptibility    
\bea
\chi^{+-}_{\alpha \beta, \mu \nu}(\q,\q^{\prime},i\omega_n)  
= \nonumber\\
T\int^{1/T}_0{d\tau e^{i \omega_{n}\tau}\langle T_\xi
[{S}^{+}_{\alpha \beta}(\q, \tau) { S}^{-}_{\nu \mu} (-\q^{\prime}, 0)]\rangle}. 
\eea
within the two-orbital model. Here, $\q, \q^{\prime} = \q$ or $\q + \Q$. The components of the spin operators
are given by $
{ S}^{i}_{\alpha \beta} (\q)= \sum_{\bf k} \sum_{\sigma \sigma^{\prime}}
d^{\dagger}_{\alpha \sigma}(\k+\q) \sigma^{i}_{\sigma \sigma^{\prime}} 
d_{\beta \sigma^{\prime}}(\k)$. $\sigma^{i}$ are Pauli matrices while 
the subscripts $\sigma, \sigma^{\prime} = \uparrow, \downarrow$. 
Using the mean-field Hamiltonian described by Eq. 4, the bare Green's 
functions $G^{\uparrow}_{\alpha \mu}(\k, i\omega_n)$ can be obtained. 
The transverse-spin susceptibility for the AFM state in the random-phase
approximation is calculated as
\be
\hat{\bar{\chi}}_{}(\q, i\omega_n) = (\hat{{\bf 1}} -
\hat{{\chi}}(\q, i\omega_n)\hat{U})^{-1} \hat{{\chi}}(\q, i\omega_n),
\ee
 where $\hat{{\bf 1}}$ is a $8 \times 8$ identity matrix and $\hat{U}$ is a 
 block-diagonal interaction matrix with the elements of both the blocks being identical. 
 $\hat{\bar{\chi}}(\q, i\omega_n)$ and $\hat{{\chi}}(\q, i\omega_n)$ are 8 $\times$ 8 matrices,
 which is evident from the structure of  Eq. 6. Note that each element of the 
 susceptibility matrix $\hat{{\chi}}(\q, i\omega_n)$ contains 
   $\chi_{\alpha \beta, \mu \nu}(\q,\q,i\omega_n) = \sum_{\k,i\omega^{\prime}_n}
G^{\uparrow}_{\alpha \mu}(\k+\q,i\omega^{\prime}_n
+i\omega_n)G^{\downarrow}_{\nu \beta} (\k,i\omega^{\prime}_n)$ when $\q^{\prime} = \q {}$ as well as 
the terms arising due to the Umklapp processes. The physical spin susceptibility using the appropriate 
 elements of the $\hat{\bar{\chi}}_{}(\q, i\omega_n) $ is given by 
 $ \bar{\chi}_{ph}(\q, i\omega_n) = \sum_{\alpha \mu} {\bar{\chi}}_{
\alpha \alpha, \mu \mu} (\q, \q, i\omega_n)$~\cite{knolle}. 

\begin{figure}[t]
\begin{center}
\vspace{-0mm}
\hspace{-7mm}
\psfig{figure=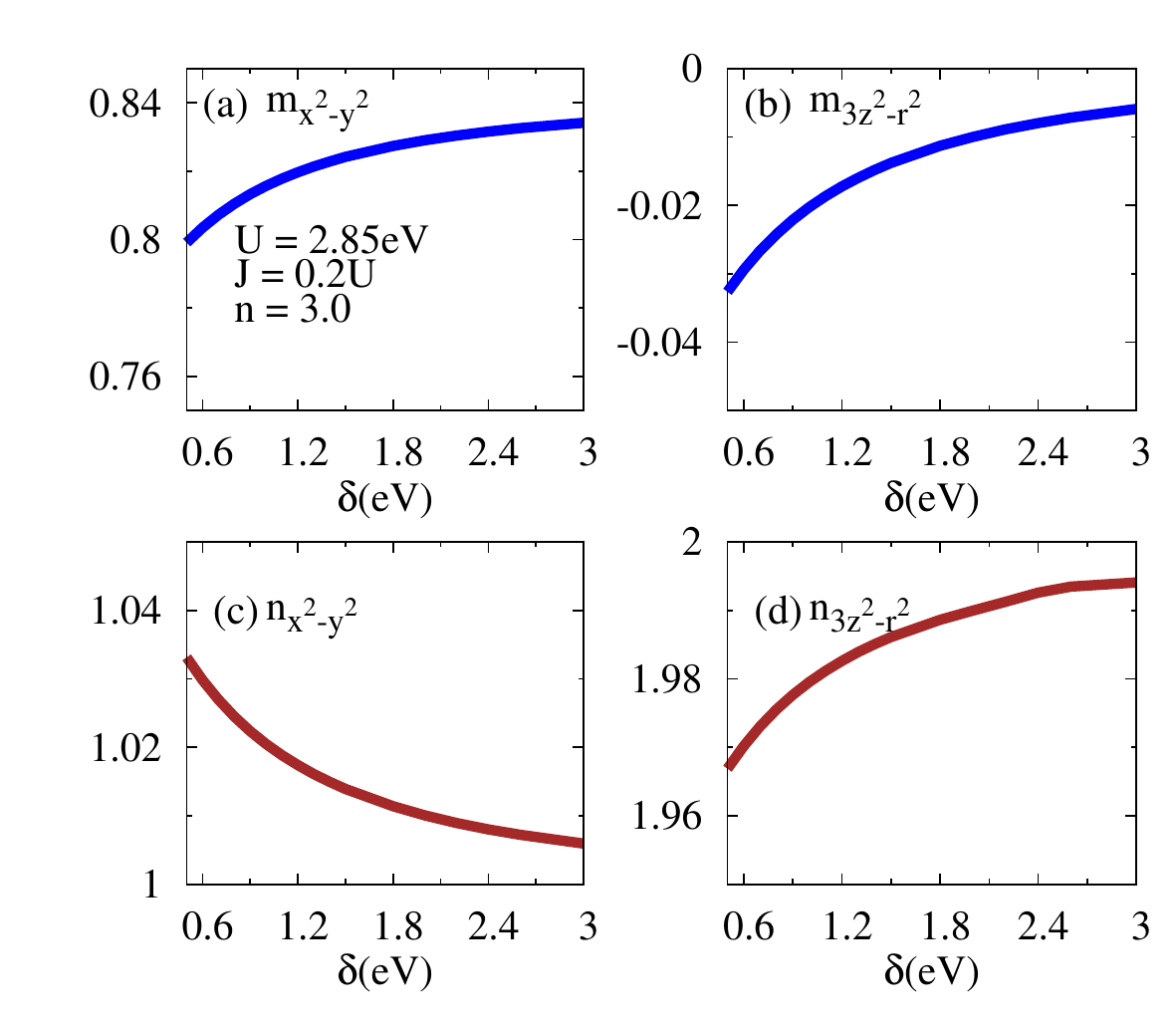,width=1.05 \linewidth}
\end{center}
\vspace{-6mm}
\caption{Magnetizations for the orbitals (a) $d_{x^2-y^2}$ 
and (b) $d_{3z^2-r^2}$ show opposite behavior as a function of the 
OS. It increases for the former while decreases in magnitude 
for the latter when the OS is increased. Note that a negative $m_{3z^2-r^2}$ indicates that 
the magnetization is oriented in a direction opposite to $m_{x^2-y^2}$. At the same time, the orbital-resolved charge density (a) 
$n_{x^2-y^2}$ decreases and (b) $n_{3z^2-r^2}$ increases.}
\label{ordparam}
\end{figure}

\section{Results and Discussion}
Fig. \ref{band} shows the electronic dispersions in the 
two-orbital model for the OS (a) $\delta = 0.5$eV 
and (b) 1.5eV. There is no mixing of $d_{x^2-y^2}$ and $d_{3z^2-r^2}$ orbitals in the 
bands along the $(0, 0)$-($\pi, \pi$) direction as $h_{12} (\k)$ vanishes 
identically. The mixing is maximum near ($\pi, 0$) in the 
vicinity of Van Hove singularity because 
$h_{12} (\k)$ attains it's maximum value at the 
same point. In other directions, the orbital mixing is 
moderate. As the OS increases, the two bands are 
increasingly orbitally polarized and they become almost 
completely polarized for $\delta \approx 1.5$eV and beyond. In the limit of very small 
$e_g$ splitting the bands will resemble to that of the monolayer manganites~\cite{dheeraj1}.

Fig. \ref{ordparam} shows the charge densities and magnetic order parameters
for the orbitals $d_{x^2-y^2}$ and $d_{3z^2-r^2}$ as 
a function of OS in the AFM state. The total charge density $n = 3.0$ is
fixed throughout the paper unless stated otherwise, which corresponds to the scenario with nearly
half-filled $d_{x^2-y^2}$ orbital and completely filled $d_{3z^2-r^2}$ orbital. The charge density $n_{x^2-y^2}$ 
in the $d_{x^2-y^2}$ orbital decreases while
$n_{3z^2-r^2}$ increases as the OS increases, which results from the constraint that the 
total charge density is fixed while the electrons will occupy the low-energy states first. However, $n_{x^2-y^2} > 1$ by $\approx 2\%$ even if  
$\delta \approx 1.5$eV a value greater than what is considered widely acceptable for LSCO. Thus, the $d_{3z^2-r^2}$ orbital 
is not completely filled and therefore can play an important role in the spin-wave excitations to be discussed below. 
The magnetization $m_{x^2-y^2}$ in the $d_{x^2-y^2}$ orbital increases continuously with a rise in the OS. 
This is mainly a consequence of the fact that the double occupancy diminishes as the OS increases. On the other hand,
the magnitude of $m_{3z^2-r^2}$ drops as the double occupancy increases.
\begin{figure}[t]
\begin{center}
\vspace{0mm}
\hspace{-0mm}
\psfig{figure=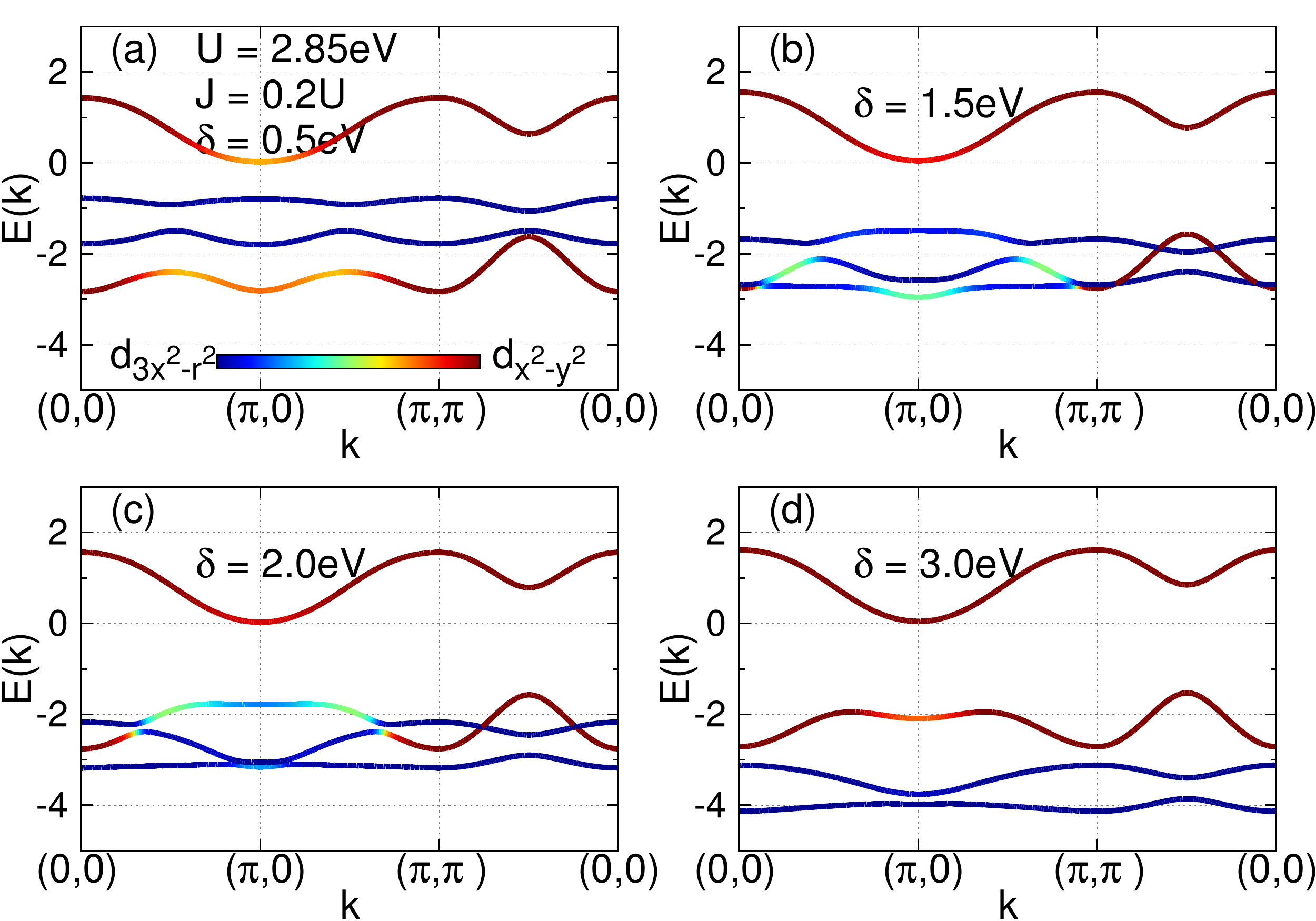,width=0.7 \linewidth, angle = -90}
\end{center}
\vspace{-3mm}
\caption{Predominant orbitals in all the four reconstructed bands of the AFM state along the high-symmetry directions for 
$\delta$ = (a) 0.5eV, (b) 1.5eV, (c) 2.0eV and (d) 3.0eV. A significant orbital mixing in the lower exchange-split band 
and $d_{3z^2-r^2}$ dominated band is present for $\delta \approx 2$eV.}
\vspace{0mm}
\label{magband}
\end{figure}

Fig. \ref{magband} shows the reconstructed band in
the AFM state for various OSs. Unlike the $d_{x^2-y^2}$ band, the exchange 
splitting for $d_{3z^2-r^2}$ dominated band is small because $d_{3z^2-r^2}$ orbitals are 
nearly doubly occupied. For $\delta = 0.5$eV, the 
$d_{3z^2-r^2}$ dominated bands are located in between the exchange-split 
$d_{x^2-y^2}$ bands. Moreover, the mixing of orbitals in the bands are minimal. 
But the separation between the relatively narrow $d_{3z^2-r^2}$ 
band and the lower-exchange split $d_{x^2-y^2}$ band decreases with the rise in OS. This results in an increased mixing 
of the two orbitals in the three low lying bands. With further
rise in the OS, the orbital mixing maximizes and thereafter it decreases so that there are
almost completely polarized two upper bands dominated by the $d_{x^2-y^2}$ orbital and 
two lower bands dominated by the $d_{3z^2-r^2}$ orbital as shown in Fig.~\ref{magband}(d).

Fig.~\ref{spinwave} shows the spin-wave excitation energy calculated by using imaginary part of 
$ \bar{\chi}_{ph}(\q, i\omega_n)$ as a function of $\delta$ along 
the high-symmetry direction. We have chosen the intra-orbital Coulomb interaction parameter $U = 2.85$eV, Hund's
coupling $J = 0.2U$ and  $\delta \approx$ 1eV so that the spin-wave excitations shows a good 
agreement with the neutron-scattering experiments for LSCO. It is worthwhile to note that there may a 
non-negligible magnon-self energy correction due to coupling of spin degree of freedom with charge and orbital degree of freedom~\cite{dheeraj2}. 

Earlier, the necessity of a similar range of $U$ was stressed in the one-orbital model for
different cuprates~\cite{coldea, delannoy,avinash}. The estimates by studies based on the
LDA + DMFT or the photoemission spectroscopy also yields a similar value of $U$ $\sim$ 3eV~\cite{jang, nilsson}. For this 
range of on-site Coulomb interaction parameter, the magnetization
$m_{x^2-y^2} \sim 0.9$ in a self-consistent meanfield theory, which is significantly larger than 
the experimental estimates ~0.55$\mu_B$. However, by going beyond the meanfield level,
it can be shown that the correction to the meanfield sublattice magnetization originating due to 
the spin fluctuations may yield a reduction up to 40\%~\cite{avinash1}. This brings the sublattice 
magnetization to a value very close to what is observed experimentally. It is true that the corrections to the 
sublattice magnetization were obtained only within the one-orbital model, it is though not unreasonable 
to expect that the magnitude of correction will be of a similar order even 
in the two-orbital model.

More importantly, the zone-boundary excitations show a significant dependence on the OS. 
In particular, we find that the zone-boundary spin-wave energy increases with the OS within 
the range $0.5$eV $\lesssim \delta \lesssim$ 2.0eV. The growth is monotonic at the high 
symmetry point (1/2, 0). Beyond $\delta \sim$ 2.0eV, the zone-boundary spin-wave energy starts decreasing but the 
rate of decline is comparatively smaller than the rate of rise noted for $\delta \lesssim $2.0eV.

The hardening of zone-boundary spin-wave energy also implies an enhancement in the effective exchange coupling, which in 
turn may indicate a more stable AFM state. The stabilization may result from the presence of an additional 
channel for lowering of energy, which is specific to the two-orbital models. While in the one-orbital model, 
there is only a single channel for the second order super-exchange interaction, several channels for the second 
order exchange interactions are possible in the two-orbital models.

\begin{figure}[t]
\begin{center}
\hspace{-2mm}
\psfig{figure=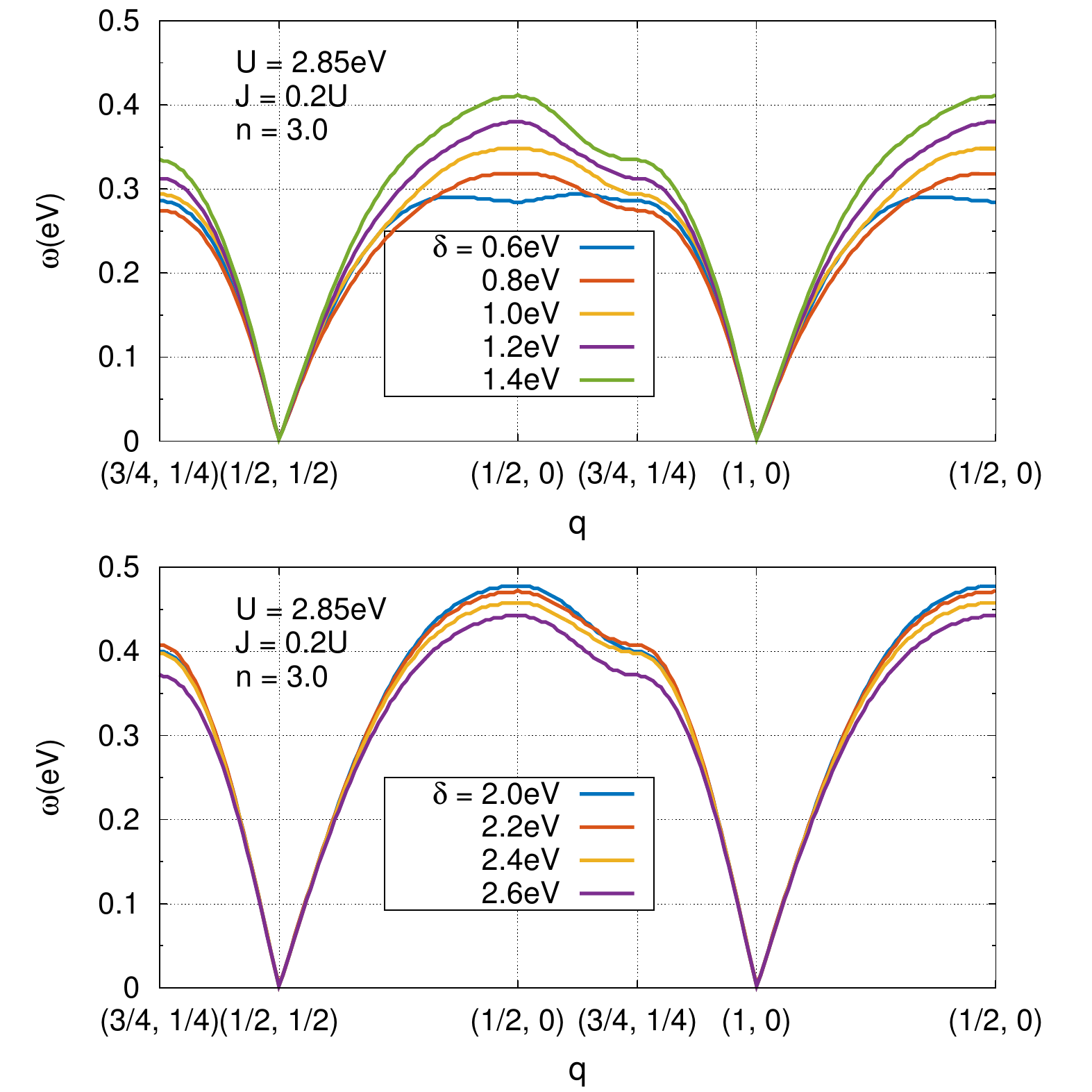,width=1.07 \linewidth}
\end{center}
\vspace{-5mm}
\caption{Spin-wave excitations along the high-symmetry directions. Zone-boundary spin-wave energy shows a 
significant dependence on the orbital splitting between $d_{x^2-y^2}$ and $d_{3z^2-r^2}$ orbitals 
within the range $0.5$eV$\lesssim\delta\lesssim$2.0eV, while there is 
only a weaker dependence beyond $\delta=$2.0eV.}
\vspace{-2mm}
\label{spinwave}
\end{figure}
The zone-boundary hardening of the spin-wave excitation energy occurs because of the virtual process involving 
the hopping of a $d_{3z^2-r^2}$ electron to a neighboring $d_{x^2-y^2}$ orbital and then returning back, 
which is possible as the inter-orbital hopping is non zero. More specifically, when the 
magnetic moments are considered oriented along the $z$-direction as in the current paper, an additional 
antiferromagnetic exchange coupling can be generated because an $\uparrow$-spin $d_{3z^2-r^2}$ electron from a site with $\uparrow$-spin $d_{x^2-y^2}$ electron 
can hop to the $d_{x^2-y^2}$ orbital at a nearest-neighbor site occupied already by a $\downarrow$-spin electron and then 
return back to it's original position.

The energy of the $\uparrow$-spin $d_{3z^2-r^2}$ electron at it's original site is $\approx U+U^{\prime}-J$.
The $d_{3z^2-r^2}$ orbital is doubly occupied therefore there is a contribution of $U$
due to the intra-orbital Coulomb interaction. Similarly, there is also a contribution of $U^{\prime}$
due to the inter-orbital Coulomb interaction. Further, the energy is lowered by $J$ because of 
the Hund's coupling between the $\uparrow$-spin $d_{3z^2-r^2}$ and $d_{x^2-y^2}$ electrons at the  
original site.

When the $\uparrow$-spin $d_{3z^2-r^2}$ electron transfers to the $d_{x^2-y^2}$ orbital 
at the neighboring site, it's Coulombic energies are $U$ and $2U^{\prime}$ due to the 
intra- and inter-orbital interactions, respectively. There is no contribution from the 
Hund's coupling term because $d_{3z^2-r^2}$ is almost doubly occupied. Thus, the total 
Coulombic energy at the new site is $\approx U+2U^{\prime}+\delta$, where we have also incorporated 
the fact that $d_{x^2-y^2}$ and $d_{3z^2-r^2}$ orbitals are separated by the energy $\delta$ because of the 
orbital splitting.

Such a process leads to the antiferromagnetic exchange 
coupling $J_{int} \approx -4t^2_4/(\delta+ U^{\prime}+J)= -4t^2_4/(\delta + U -J)$, where $t_4$ is the 
inter-orbital hopping parameter and $U = U^{\prime} + 2J$ in accordance with the rotational symmetry of the Hamiltonian. The contribution of this term is negligible 
when the lower exchange-split $d_{x^2-y^2}$ band and $d_{3z^2-r^2}$ band don't mix much as is the case when the 
OS is either very small ($\sim$ 0.5eV) or too large ($\sim$ 3.0eV). However, at the intermediate value of 
$\delta \sim$ 2eV, the contribution is significant as the orbital mixing is non-negligible. Note the presence of $\delta$ 
in the denominator of $J_{int}$, which explains a faster rise in the zone-boundary spin-wave energy for a 
smaller OS within the range $0.5$eV $\lesssim \delta \lesssim$ 2.0eV and a slower decline beyond $\delta \sim$ 2.0eV 
for a larger OS.

\begin{figure}[]
\begin{center}
\vspace{0mm}
\hspace{-2mm}
\psfig{figure=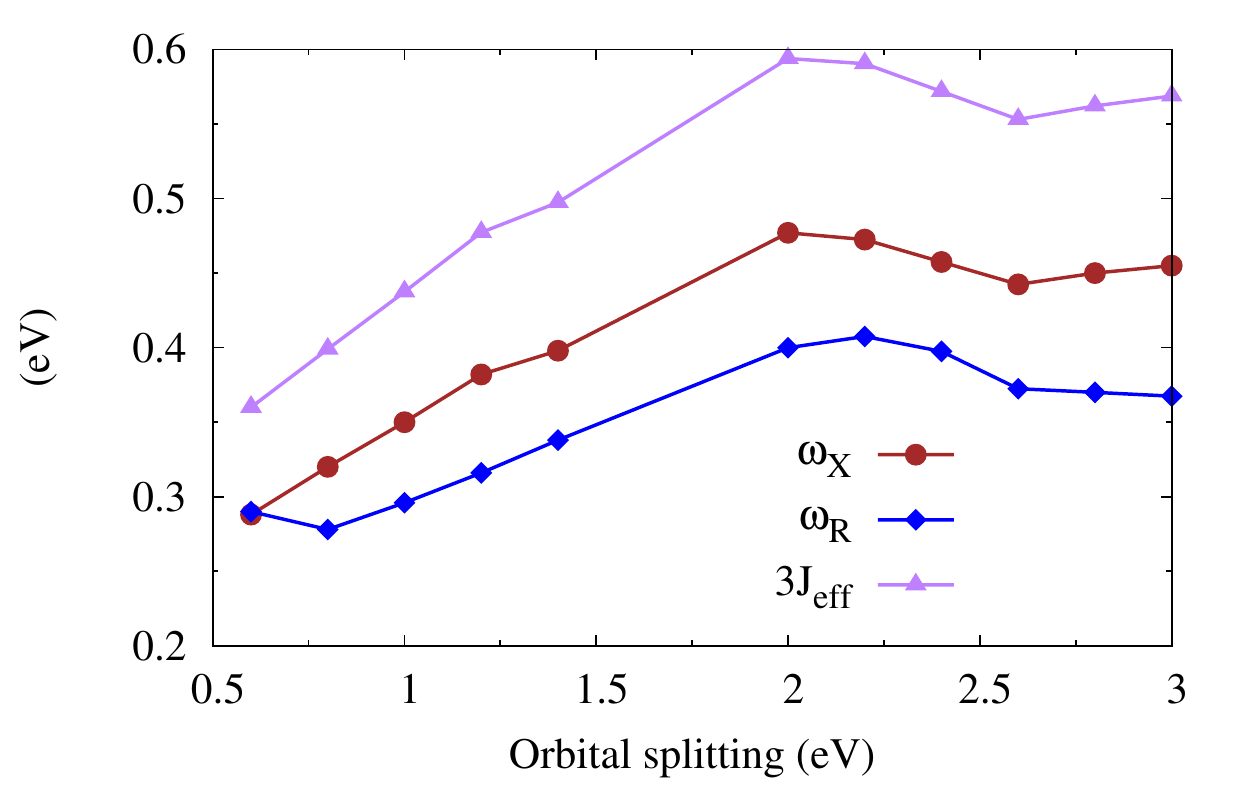,width=1.05 \linewidth,angle = 0}
\end{center}
\vspace{-5mm}
\caption{The zone-boundary spin-wave excitation energies and the effective exchange coupling as a function 
of orbital splitting between the $d_{x^2-y^2}$ and $d_{3z^2-r^2}$ orbitals.}
\vspace{-4mm}
\label{exch}
\end{figure}
Fig. \ref{exch} shows the zone-boundary spin-wave excitations $\omega$ at X $\equiv$ (1/2, 0) and R $\equiv$ (3/4, 1/4) as 
a function of $\delta$. Both show almost a linear growth up to $\delta \approx 2.0$eV, thereafter they 
don't show much variation. The region with a nearly linear dependence shows a very good qualitative agreement 
with the observed zone-boundary spin-wave excitations in the samples subjected to the in-plane strain. We have also plotted 
the effective-exchange coupling determined from the approximate relation $J_{eff} = \omega_X/2Z$, where the renormalization factor resulting 
from the self-energy correction is $Z = 1.2$~\cite{delannoy,ivashko1}. As the exchange coupling displays a 
nearly linear dependence on the in-plane compressive strain, our finding suggests that the $e_g$ orbital 
splitting $\delta$ may also exhibit a nearly linear dependence on the in-plane strain.

\section{Summary and conclusions}
The spin-wave excitations in the cuprates 
has been largely explored within a one-orbital model. 
In a recent work, the two-orbital model was invoked to 
explain the difference between the superconducting-transition 
temperature across different cuprates using the fluctuation-exchange approximation which
incorporates the spin-spin correlations~\cite{sakakibara,sakakibara1,sakakibara2,tang}. An indirect implication of the above
result is that the spin-wave excitations in different phases including the AFM
 may also show dependence on the OS induced by the in-plain stress. 

The strain in the layered cuprates affects not only the overlap integral 
between the orbitals at neighboring sites but it can also lead to a non negligible 
modification in the on site Coulombic interaction. The net possible impact of the interplay between the aforementioned 
consequences on the spin-wave excitations 
is yet to be fully understood. However, the most significant impact of in-plain strain perhaps is
on the extent of orbital mixing for the bands either located near or far from the Fermi surface. 
As illustrated through the current work, even if we ignore the modification in 
overlap integral and Coulomb interaction, the two-orbital model successfully 
describes the experimental observations in terms of orbital mixing present in various bands.

Our study is focused at zero doping where the cuprates show only 
the antiferromagnetic order. On doping holes, the long-range magnetic order is lost. However, 
the nature of leading order local magnetic-exchange couplings are expected to show a weak dependence on 
doping. The higher spin-wave excitation energy for a larger $\delta$ may indicate an 
enhancement in the exchange coupling which will remain true even on doping holes. As found, 
the spin-wave excitation energy increases with $\delta$ for a realistic range so does the magnetic-exchange coupling. Therefore, the high 
energy spin-fluctuations would help to increase the superconducting transition temperature $T_c$. However, this is true only for those cuprates for which $\delta \lesssim 2$eV. The cuprates
such as HgBa$_2$CuO$_4$ has a relatively larger $\delta \gtrsim$  2eV. According to 
our calculation, the spin-wave excitation energy does not increase on increasing $\delta$ near 2eV. Thus, 
by applying in-plane compressive strain we may not be able to increase $T_C$ of HgBa$_2$CuO$_4$ 
except for the lower $e_g$ split cuprates such as 
La$_2$CuO$_4$~\cite{sakakibara,sakakibara1,sakakibara2}.

In summary, we have explored the spin-wave excitations in the undoped AFM state of cuprates within 
a two-orbital model based on $d_{x^2-y^2}$ and $d_{3z^2-r^2}$ orbitals. Our investigation reveals 
that the zone-boundary spin-wave energy hardens with an increase in the orbital splitting for
the range $0.5$eV $\lesssim \delta \lesssim$ 2.0eV. The result, besides providing a plausible explanation 
for the recent observations in RIXS measurements, emphasizes also on the importance of 
$d_{3z^2-r^2}$ orbital in the cuprates with smaller $e_g$ splitting.
\section*{Acknowledgement}
We would like to thank the anonymous referee for bringing our attention to the issue explored in this 
work. We acknowledge the use of HPC clusters at HRI. D. K. Singh was supported through 
start-up grant SRG/2020/002144 funded by DST-SERB. Y. Bang was supported through NRF Grant No. 
2020-R1A2C2-007930 funded by the National Research Foundation of Korea.

\end{document}